\documentclass[prd,superscriptaddress,amsfonts,amssymb,amsmath,showpacs,twocolumn,nofootinbib]{revtex4-2}
\usepackage{bm}
\usepackage{amsfonts}
\usepackage{latexsym}
\usepackage{graphicx}
\usepackage{amsmath}
\usepackage{palatino}
\usepackage{xcolor} 
\usepackage{mathpazo}
\usepackage{xcolor}
\usepackage{rotating}
\usepackage{fontawesome5} 
\usepackage{adjustbox}
\usepackage{tensor}
\usepackage{textcomp}
\usepackage{float}
\usepackage{booktabs}
\usepackage{dcolumn}
\usepackage[titletoc]{appendix}
\usepackage{booktabs}
\usepackage{multirow}
\usepackage{hyperref}
\hypersetup{colorlinks,citecolor=blue}
\usepackage{amsmath}
\usepackage{xcolor}
\usepackage{orcidlink}
\usepackage{epsfig}
\usepackage{caption}
\usepackage{subcaption}
\usepackage{commath}

\hypersetup{colorlinks,citecolor=blue}
\hypersetup{colorlinks=true,linkcolor=magenta,filecolor=magenta,    urlcolor=blue}

\def\be{\begin{equation}}
\def\ee{\end{equation}}
\def\bea{\begin{eqnarray}}
\def\eea{\end{eqnarray}}

\begin{document}

\title{Conformal Killing Gravity: New Constraints from DESI DR2 BAO datasets}

\author{Himanshu Chaudhary}
\email{himanshu.chaudhary@ubbcluj.ro,\\
himanshuch1729@gmail.com}
\affiliation{Department of Physics, Babeș-Bolyai University, Kogălniceanu Street, Cluj-Napoca, 400084, Romania}
\affiliation{Research Center of Astrophysics and Cosmology, Khazar University, Baku, AZ1096, 41 Mehseti Street, Azerbaijan}
\author{Salvatore Capozziello}
\email{capozziello@na.infn.it}
\affiliation{Dipartimento di Fisica ``E. Pancini", Universit\`a di Napoli ``Federico II", Complesso Universitario di Monte Sant’ Angelo, Edificio G, Via Cinthia, I-80126, Napoli, Italy,}
\affiliation{Istituto Nazionale di Fisica Nucleare (INFN), sez. di Napoli, Via Cinthia 9, I-80126 Napoli, Italy,}
\affiliation{Scuola Superiore Meridionale, Largo S. Marcellino, I-80138, Napoli, Italy.}
\author{Carlo Alberto Mantica}
\email{carlo.mantica@mi.infn.it}
\affiliation{Dipartimento di Fisica  Aldo Pontremoli, Universit\`a degli Studi di Milano and INFN,  Sez, di Milano, Via Celoria 16, 20133 Milano, Italy}
\author{Luca Guido Molinari}
\email{luca.molinari@mi.infn.it}
\affiliation{Dipartimento di Fisica  Aldo Pontremoli, Universit\`a degli Studi di Milano and INFN,  Sez, di Milano, Via Celoria 16, 20133 Milano, Italy}

\author{Dhruba Jyoti Gogoi}
\email{moloydhruba@yahoo.in}
\affiliation{Department of Physics, Madhabdev University, Narayanpur, Lakhimpur 784164, Assam, India}

\author{G. Mustafa}
\email{gmustafa3828@gmail.com}
\affiliation{Department of Physics,
Zhejiang Normal University, Jinhua 321004, People’s Republic of China}

\begin{abstract}
We investigate a geometric approach referred to as the Conformal Killing Gravity (CKG), in which the dark energy sector emerges naturally from the conformal Killing symmetry of the Robertson--Walker space-time. Within this framework, the divergence-free conformal Killing tensor behaves as an effective perfect fluid, giving rise to a dynamical dark-energy component whose density, pressure, and equation of state are uniquely determined by the underlying geometry, without introducing any empirical dark-energy parametrization. The resulting CKG model extends the standard $\Lambda$CDM cosmology through a single additional parameter while recovering the $\Lambda$CDM limit in the absence of the geometric contribution. We constrain the model using Planck PR4 (NPIPE) CMB temperature, polarization, and lensing observations, ACT DR6 CMB lensing, DESI DR2 baryon acoustic oscillation measurements, and the Pantheon$+$, DES-Dovekie, and Union3 Type Ia supernova compilations. The analysis shows that the CKG  favors a quintessence-like dark-energy evolution with no evidence for phantom crossing, while the reconstructed equation of state rapidly approaches the cosmological constant at earlier cosmic times. Furthermore, the model predicts a future critical redshift in the range $-0.8 \lesssim z_c \lesssim -0.7$, indicating that the present cosmic expansion eventually reaches a turning point before the formal singular limit at $z=-1$. Since the geometric contribution modifies only the post-recombination expansion history, the sound horizon remains essentially unchanged and the model does not provide a complete solution to the $H_0$ tension. Nevertheless, it remains fully consistent with current weak-lensing constraints on $S_8$ and yields strong Bayesian evidence over the standard $\Lambda$CDM model. Our results demonstrate that the CKG  provides a simple, physically motivated, and observationally favored framework for describing the late-time accelerated expansion of the Universe.
\end{abstract}. 


\maketitle

\section{Introduction}\label{sec_1}
The concordance $\Lambda$CDM model currently serves as the foundational framework of modern cosmology, yielding robust agreements with observations of the cosmic microwave background (CMB) anisotropies, large-scale structure (LSS) formation, and the late-time accelerated expansion of the Universe \cite{Planck:2018vyg, SupernovaSearchTeam:1998fmf,SupernovaCosmologyProject:1998vns,Padmanabhan:2002ji,Copeland:2006wr}. Nevertheless, this paradigm is hindered by persistent theoretical difficulties, most notably the fine-tuning and cosmic coincidence problems associated with the cosmological constant \cite{Weinberg:1988cp,Velten:2014nra}. Compounding these theoretical shortcomings are mounting observational anomalies. Primary among these are the $H_0$ and $S_8$ tensions, which indicate statistically significant discrepancies between parameter inferences from early-Universe phenomena and those from direct, late-time local probes \cite{DiValentino:2021izs,Perivolaropoulos:2021jda,Abdalla:2022yfr}.

The landscape of precision cosmology has recently been catalyzed by the Dark Energy Spectroscopic Instrument (DESI) Data Release 2 (DR2) \cite{abdul2025desi}. These new Baryon Acoustic Oscillation (BAO) measurements show a tantalizing preference for a dynamical dark energy equation of state over a static cosmological constant, driving the community to reinvestigate phenomenological and geometric alternatives to standard General Relativity (GR). A compelling theoretical alternative is the Conformal Killing Gravity (CKG). Originally introduced by Harada to address the dark energy puzzle without invoking exotic matter \cite{Harada:2023rqw, Harada:2023afu}, CKG was later reformulated in Ref. \cite{mantica2023note}. This reformulation demonstrated that CKG is equivalent to standard Einstein equations supplemented by a divergence-free conformal Killing tensor. This tensor acts as a purely geometric source term, effectively behaving as a dynamical dark fluid. In the cosmological context, this introduces a novel density parameter, $\Omega_D$, which uniquely governs the post-recombination expansion history of the Universe \cite{mantica2023note}.

In this paper, we advance the investigation of CKG cosmology by performing a robust, high-precision Markov Chain Monte Carlo (MCMC) analysis. We utilize the most up-to-date and complementary cosmological datasets, combining the newly released DESI DR2 BAO measurements with the state-of-the-art \texttt{CamSpec} CMB likelihoods from the Planck \texttt{NPIPE PR4} release \cite{Efstathiou2021Detailed,rosenberg2022cmb}, alongside ACT DR6 CMB lensing data \cite{madhavacheril2024atacama,qu2024atacama}. To rigorously test the late-time expansion history and break parameter degeneracies, we systematically combine these probes with three independent Type Ia Supernovae (SNe Ia) catalogs: the Pantheon$+$ compilation \cite{scolnic2022pantheon,brout2022pantheon}, the recalibrated DES-Dovekie sample \cite{popovic2025dark}, and the Union3 compilation \cite{rubin2025union}. We improve the approach already followed in Ref.\cite{Capozziello:2025kws} fixing more reliable constraints.

Our primary objective is to evaluate the extent to which the CKG model can alleviate the $H_0$ and $S_8$ tensions while remaining consistent with the combined observational data. Furthermore, we aim to definitively test the dynamical behavior of this geometric dark energy component—specifically assessing whether the $\Omega_D$ parameter favors a quintessence regime or exhibits phantom-crossing dynamics, such as the Quintom-B scenario \cite{Cai:2025mas}. Finally, we assess the statistical viability of the CKG model against the standard $\Lambda$CDM paradigm using Bayesian evidence.

The paper is organized as follows: in Section \ref{sec_2}, we provide an overview of the theoretical background and the Friedmann equations in Conformal Killing Gravity. Section \ref{sec_3} details the recent observational datasets and the MCMC methodology utilized in this work. In Section \ref{sec_4}, we present the constraints obtained on the cosmological parameters, discuss the implications for the $H_0$ and $S_8$ tensions, and analyze the dynamical behavior of the dark energy equation of state $w(z)$. Finally, we summarize our conclusions and findings in Section \ref{sec_5}.

\section{Theoretical background}\label{sec_2}
In this section, we mainly follow Ref. \cite{Capozziello:2025kws} for summarizing the theoretical background.
The cosmological evolution considered in this work is described within the homogeneous and isotropic Robertson--Walker (RW) space-time, whose metric is

\begin{align}
ds^2 = - dt^2 + a^2(t) \left [ \frac{dr^2}{1-kr^2}+ r^2 ( d \theta^2 +\sin^2\theta d \phi^2) \right ] \label{eq:1}
\end{align}

where $a(t)$ denotes the cosmic scale factor, while $k=0,\pm1$ specifies the spatial curvature of the Universe. In the present analysis we focus on the spatially flat case, although we first introduce the general theoretical formulation for completeness.

An equivalent covariant description of the RW geometry can be established through a normalized timelike vector field satisfying $u_a u^a=-1$. Within this framework, the space-time admits the following geometric relations (see \cite{capozziello2022geometric})

\begin{align}
\nabla_{b}u_{a}
&=
H(g_{ab}+u_{b}u_{a}),
\label{eq:torse-forming}
\\
\nabla_bH
&=
-\dot Hu_b ,
\label{DOTH}
\end{align}
where the Weyl tensor identically vanishes. In comoving coordinates associated with Eq.~(\ref{eq:1}), the Hubble expansion rate is  $H=\dot a/a$, with $u^0=-1$ and $u^\mu=0$. Throughout this work, an overdot denotes the derivative along the four-velocity,
$
\dot{f}=u^a\nabla_af=\partial_tf,
$
which corresponds to differentiation with respect to cosmic time.

Eq.~(\ref{DOTH})  implies that the four-velocity is an eigenvector of the Ricci tensor, $R_{ab}u^b=\xi u_a,$ where the corresponding eigenvalue is directly related to the cosmic acceleration through

\begin{align}
\xi
=
3(H^2+\dot H)
=
3\frac{\ddot a}{a}.
\label{xiACC}
\end{align}

For the RW geometry, both the Ricci tensor and the Ricci scalar can be expressed as

\begin{align}
R_{ab}
&=
\frac{R-4\xi}{3}u_au_b
+
\frac{R-\xi}{3}g_{ab},
\label{eq:2.3 Ricci GRW}
\\
R
&=
\frac{R^{\star}}{a^{2}}
+
6H^{2}
+
2\xi,
\label{eq:2.9 scalar R}
\end{align}
where $R^{\star}=6k$ denotes the intrinsic curvature of the spatial hypersurfaces.

A characteristic ingredient of CKG is the existence of a divergence-free conformal Killing tensor. For the RW background, this tensor assumes the form

\begin{align}
K_{ab}
=
g_{ab}\left[
\frac56Ca^2-\Lambda
\right]
+
u_au_b\frac{Ca^2}{3},
\label{eq:Conformal Killing GRW}
\end{align}

where $C$ and $\Lambda$ arise as integration constants \cite{mantica2023note,mantica2024conformal}. The expression above follows directly from the conformal Killing symmetry of the RW space-time. In particular, the geometry admits the unique timelike conformal Killing vector $\xi_a=a(t)u_a$, satisfying

\begin{equation}
\nabla_a\xi_b+\nabla_b\xi_a
=
2\dot ag_{ab}.
\end{equation}

Consequently, the tensor $K_{ab}
=
\frac13C\xi_a\xi_b
+
\beta(t)g_{ab}$ is itself a conformal Killing tensor, where $C$ is an arbitrary constant and $\beta(t)$ is an unknown function. Requiring this tensor to be divergence-free leads to $\dot\beta=\frac53a\dot a,$ whose integration gives $\beta(t)
=
\frac56Ca^2(t)-\Lambda.$ 

Substituting this result into the previous expression immediately recovers Eq.~(\ref{eq:Conformal Killing GRW}). 

The conformal Killing tensor can be interpreted as an effective perfect fluid of purely geometric origin, $K_{ab}
=
(p_D+\mu_D)u_au_b+p_Dg_{ab},$ where the corresponding effective dark energy density and pressure are 

\begin{align}
\mu_D
&=
-\frac12Ca^2+\Lambda,
\label{MUD}
\\
p_D
&=
\frac56Ca^2-\Lambda.
\label{PD}
\end{align}

Unlike the cosmological constant of the standard $\Lambda$CDM scenario, both quantities evolve explicitly with the scale factor, reflecting the dynamical nature of the geometric dark sector in CKG cosmology. The associated dark energy equation-of-state parameter therefore becomes

\begin{equation}
w_D(a)
=
\frac{p_D}{\mu_D}
=
-1+
\frac{2Ca^2}
{6\Lambda-3Ca^2},
\label{eq:w(a)}
\end{equation}
which follows directly from the field equations of the theory. In contrast to phenomenological parameterizations, such as the Chevallier--Polarski--Linder \cite{chevallier2001accelerating,linder2003exploring} or the Jassal--Bagla--Padmanabhan (JBP) model \cite{jassal2005wmap}, this evolution is not assumed a priori but instead emerges naturally from the underlying conformal Killing symmetry.

The effective energy--momentum tensor describing the ordinary matter sector is obtained by combining Eqs.~(\ref{eq:2.3 Ricci GRW}) and (\ref{eq:Conformal Killing GRW}), yielding

\begin{align}
T_{ab}
&=
R_{ab}
-\frac12Rg_{ab}
-K_{ab}
\nonumber\\
&=
-\frac16\left(R+2\xi+5Ca^2-6\Lambda\right)g_{ab}
\nonumber\\
&\quad
+\frac13\left(R-4\xi-Ca^2\right)u_au_b
\nonumber\\
&\equiv
(p+\mu)u_au_b
+
pg_{ab},
\label{eq:stress energy conformal killing}
\end{align}

where $\mu$ and $p$ denote the energy density and pressure of the ordinary matter component, respectively. Substituting Eqs.~(\ref{eq:2.9 scalar R}) and (\ref{xiACC}) into the above expression leads to the modified Friedmann equations,

\begin{align}
\mu
&=
\frac{R^\star}{2a^2}
+
3H^2
+
\frac12Ca^2
-
\Lambda,
\label{KAPPAMU}
\\
p
&=
-\frac{R^\star}{6a^2}
-
3H^2
-
2\dot H
-
\frac56Ca^2
+
\Lambda.
\label{eq:Friedmann conformal killing}
\end{align}

It is straightforward to verify that the standard Einstein equations with a cosmological constant are recovered in the limit $C=0$, corresponding to the $\Lambda$CDM model.

We assume that the ordinary matter sector consists of pressureless matter and radiation, with $p_R=\mu_R/3$. Conservation of the matter and radiation fluids implies

\[
\frac{\mu_M}{a^3}
=
\frac{\mu_{M0}}{a_0^3},
\qquad
\frac{\mu_R}{a^4}
=
\frac{\mu_{R0}}{a_0^4}.
\]

Substituting these conservation laws into Eq.~(\ref{KAPPAMU}) gives the modified Friedmann equation,

\begin{equation}
\mu_{M0}\left(\frac{a}{a_0}\right)^{-3}
+
\mu_{R0}\left(\frac{a}{a_0}\right)^{-4}
=
\frac{R^\star}{2a^2}
+
3H^2
+
\frac12Ca^2
-
\Lambda.
\end{equation}

Dividing the above equation by $3H_0^2$ and introducing the dimensionless density parameters yields

\begin{align}
\left(\frac{H}{H_0}\right)^2
=&
\Omega_R\left(\frac{a}{a_0}\right)^{-4}
+
\Omega_M\left(\frac{a}{a_0}\right)^{-3}
\nonumber\\
&
+\Omega_k\left(\frac{a}{a_0}\right)^{-2}
+\Omega_\Lambda
+\Omega_D\left(\frac{a}{a_0}\right)^2,
\label{eq:ALL H}
\end{align}

where

\begin{gather}
\Omega_M=\frac{\mu_{M0}}{3H_0^2},
\qquad
\Omega_R=\frac{\mu_{R0}}{3H_0^2},
\qquad
\Omega_k=-\frac{R^\star}{6H_0^2a_0^2},
\nonumber\\
\Omega_\Lambda=\frac{\Lambda}{3H_0^2},
\qquad
\Omega_D=-\frac{Ca_0^2}{6H_0^2},
\nonumber\\
\Omega_M+\Omega_R+\Omega_k+\Omega_\Lambda+\Omega_D=1.
\end{gather}

Among these parameters, $\Omega_M$ and $\Omega_R$ correspond to the physical matter and radiation energy densities and are therefore positive. In contrast, $\Omega_\Lambda$, $\Omega_k$, and $\Omega_D$ arise from the geometric sector of the theory, and consequently their signs are not restricted to be positive.

Expressing the expansion history in terms of the redshift, defined by $1+z=a_0/a$, the Hubble parameter becomes

\begin{align}
\left(\frac{H(z)}{H_0}\right)^2
=&
\Omega_R(1+z)^4
+
\Omega_M(1+z)^3
\nonumber\\
&
+\Omega_k(1+z)^2
+\Omega_\Lambda
+
\frac{\Omega_D}{(1+z)^2}.
\label{HSQ}
\end{align}

As discussed at the beginning of this section, our analysis is restricted to a spatially flat FLRW universe ($\Omega_k=0$). Under these assumptions, the background expansion rate reduces to

\begin{equation}
\left(\frac{H(z)}{H_{0}}\right)^{2}
=
\Omega_M(1+z)^{3}
+
\Omega_{\Lambda}
+
\frac{\Omega_D}{(1+z)^{2}}\,.
\label{eq:Hubble late time}
\end{equation}

To characterize the background expansion, we introduce the cosmographic parameters through the Taylor expansion of the scale factor about the present cosmic time \cite{visser2004jerk},

\begin{align}
\frac{a(t)}{a_0}
=&
1
+
H_0(t-t_0)
-
\frac{H_0^2}{2}q_0(t-t_0)^2
\nonumber\\
&
+
\frac{H_0^3}{3!}j_0(t-t_0)^3
+
\frac{H_0^4}{4!}s_0(t-t_0)^4
+\cdots.
\end{align}

The deceleration parameter is given by $q
=
-\frac{a\ddot a}{\dot a^{\,2}}
=
-1-\frac{\dot H}{H^2}.$ Using $\dot H=(dH/dz)\dot z$ together with $\dot z=-H(1+z)$, one obtains

\begin{align}
q(z)
=
\frac{\Omega_M(1+z)^5-2\Omega_\Lambda(1+z)^2-4\Omega_D}
{2\Omega_M(1+z)^5+2\Omega_\Lambda(1+z)^2+2\Omega_D}.
\label{QZ}
\end{align}

The contraction $R_{ab}u^au^b=-\xi=-3\ddot a/a=3H^2q$ is directly related to the strong energy condition,
\begin{equation}
\left(T_{ab}-\frac{1}{2}Tg_{ab}\right)u^au^b\ge0,
\end{equation}
which, for a perfect fluid, reduces to $3p+\mu\ge0$. In the standard Einstein equations, this condition implies $3H^2q\ge0$, which is incompatible with the observed late-time accelerated expansion. Within the CKG framework, however, the gravitational source is modified to $T_{ab}+K_{ab}$, allowing accelerated expansion while preserving the geometric structure of the theory.

The weak energy condition requires the energy density measured by a comoving observer to remain non-negative. Neglecting the radiation component, this condition becomes

\begin{equation}
\Omega_M(1+z)^3+\Omega_\Lambda+\Omega_D(1+z)^{-2}\ge0,
\end{equation}
which is equivalent to the requirement $H^2(z)\ge0$ through Eq.~(\ref{eq:Hubble late time}).

Another quantity of cosmological interest is the lookback time, defined as the interval between the emission of a photon at cosmic time $t_e$ and its observation at the present epoch $t_0$. It is given by

\begin{equation}
t_{L}
=
\int_{t_e}^{t_0}dt
=
\int_{a_e}^{a_0}\frac{da}{\dot a}
=
\int_0^{z_e}\frac{dz}{H(z)(1+z)},
\label{LKBT}
\end{equation}
where the age of the Universe is obtained in the limit $z\rightarrow\infty$.

The effective dark energy density and pressure as functions of redshift are

\begin{equation}
\begin{aligned}
\mu_D
&=
3H_0^2
\left[
\Omega_\Lambda
+
\frac{\Omega_D}{(1+z)^2}
\right],
\\
p_D
&=
-3H_0^2
\left[
\Omega_\Lambda
+
\frac{5\Omega_D}{3(1+z)^2}
\right].
\end{aligned}
\label{eq:dark pressure and energy}
\end{equation}

The corresponding dark energy equation-of-state parameter is

\begin{equation}
w_D(z)
=
-1
-
\frac{2}{3}
\frac{\Omega_D}
{\Omega_D+\Omega_\Lambda(1+z)^2},
\label{eq:EoS in CKG}
\end{equation}
whose first-order expansion around the present epoch reproduces the effective CPL parametrization,

\begin{align}
w_D(z)
&=
w_{D0}
+
w_{Da}\left(\frac{z}{1+z}\right)
\nonumber\\
&=
-\frac13
\frac{5\Omega_D+3\Omega_\Lambda}
{\Omega_\Lambda+\Omega_D}
+
\frac43
\frac{\Omega_D\Omega_\Lambda}
{\left(\Omega_D+\Omega_\Lambda\right)^2}
\left(\frac{z}{1+z}\right).
\label{JBPinCKG}
\end{align}

Finally, the model predicts a limiting value at $z=-1$, where both the effective dark energy density and pressure diverge, while their ratio approaches $w_D(-1)=-5/3$. Moreover, when $\Omega_D/\Omega_\Lambda<0$, the Hubble expansion rate vanishes at a critical redshift $z_c$, defined by the condition $H(z_c)=0$. This critical redshift marks the boundary of the physically allowed cosmological evolution and is determined by

\begin{align}
\frac{\Omega_M}{\Omega_\Lambda}(1+z_c)^5
+
(1+z_c)^2
+
\frac{\Omega_D}{\Omega_\Lambda}
=
0\,.
\label{ZCRIT}
\end{align}

\section{Dataset and Methodology}\label{sec_3}
To constrain the parameters of the CKG model, we perform a (Markov Chain Monte Carlo) MCMC sampler \cite{lewis2002cosmological,lewis2013efficient,neal2005taking} using the publicly available Bayesian cosmological parameter inference package {\tt Cobaya}~\cite{torrado2021cobaya}~\href{https://github.com/CobayaSampler/cobaya}{\faGithub}. The convergence of the Markov chains is monitored using the Gelman-Rubin statistic, requiring all chains to satisfy $R-1<0.01$ before the sampling is considered converged \cite{gelman1992inference}. The chains are analyzed with the {\tt GetDist} package~\cite{lewis2025getdist}~\href{https://github.com/cmbant/getdist}{\faGithub} to extract the posteriors of each parameter, while the theoretical predictions are computed with the Boltzmann Solver Code for Anisotropies in the Microwave Background ({\tt CAMB})~\cite{lewis2000efficient,howlett2012cmb}~\href{https://github.com/cmbant/CAMB.git}{\faGithub}. 

The baseline cosmological parameter vector is chosen as $\bm{\theta}_{\Lambda\mathrm{CDM}} = \{\Omega_{\rm cdm},\,\Omega_{\rm b},\,100\,\theta_{\rm MC},\,\ln(10^{10}A_{\rm s}),\,n_{\rm s},\,\tau\}.$
For the CKG model, this parameter space is extended by introducing the additional density parameter $\Omega_D$, such that $\bm{\theta}_{\rm CKG} = \{\bm{\theta}_{\Lambda\mathrm{CDM}},\,\Omega_D\}.$ To constrain the parameters of the CKG model, we consider several combinations of early- and late-time cosmological datasets. In particular, we use the CamSpec Cosmic Microwave Background likelihood based on the Planck NPIPE PR4 release, Baryon Acoustic Oscillations, and Type Ia supernova data from Pantheon$+$, DES-Dovekie, and Union3. In the following, we briefly describe each dataset and its implementation in our analysis.

\begin{itemize}
\item \textbf{Cosmic Microwave Background:} We begin by considering the Cosmic Microwave Background (CMB) datasets. Specifically, our analysis makes use of the Planck temperature (\texttt{TT}), polarization (\texttt{EE}), and temperature--polarization cross-correlation (\texttt{TE}) power spectra, combining the low-$\ell$ \texttt{Commander} and \texttt{simall} likelihoods ($\ell<30$) with the high-$\ell$ \texttt{CamSpec} likelihood ($\ell\geq30$) from the Planck PR4 (NPIPE) release \cite{Efstathiou2021Detailed,rosenberg2022cmb}. We further incorporate the Planck PR4 CMB lensing likelihood\cite{carron2022cmb,carron2022planck}~\href{https://github.com/carronj/planck_PR4_lensing.git}{\faGithub}, along with the ACT DR6 lensing likelihood~\cite{madhavacheril2024atacama,qu2024atacama}~\href{https://github.com/ACTCollaboration/act_dr6_lenslike.git}{\faGithub}.

\item \textbf{Type Ia Supernova :} Second, we consider three Type~Ia supernova (SNe~Ia) compilations. The first is the Pantheon$+$ sample~\cite{scolnic2022pantheon}~\href{https://github.com/PantheonPlusSH0ES/DataRelease.git}{\faGithub}, which comprises 1,701 light curves from 1,550 SNe~Ia spanning the redshift range $0.001 \leq z \leq 2.26$. Following the standard Pantheon$+$ analysis, we exclude supernovae with $z<0.01$ in order to mitigate systematic uncertainties associated with peculiar velocities \cite{brout2022pantheon}. The second is the DES-Dovekie compilation~\cite{popovic2025dark}~\href{https://github.com/des-science/DES-SN5YR.git}{\faGithub}, consisting of 1,820 photometrically calibrated SNe~Ia over the redshift interval $0.02 \leq z \leq 1.14$. The third is the Union3 sample~\cite{rubin2025union}~\href{https://github.com/rubind/union3_release}{\faGithub}, containing 2,087 SNe~Ia covering the range $0.05 \leq z \leq 2.26$. These three SNe~Ia compilations offer complementary constraints on the late-time expansion history of the Universe.

\item \textbf{Baryon Acoustic Oscillation :} Finally, we consider the Baryon Acoustic Oscillation (BAO) measurements from the Dark Energy Spectroscopic Instrument (DESI) Data Release~2 (DR2)~\cite{abdul2025desi}~\href{https://github.com/CobayaSampler/bao_data.git}{\faGithub}. This dataset combines measurements from several tracers, including the Bright Galaxy Sample (BGS), Luminous Red Galaxies (LRG), Emission Line Galaxies (ELG), Quasars (QSO), and the Lyman-$\alpha$ forest, covering the redshift range $0.3 \leq z \leq 2.33$. The DESI DR2 BAO likelihood is constructed from the compressed distance measurements $D_M/r_d$, $D_H/r_d$, and $D_V/r_d$, where $D_H(z)=c/H(z)$ denotes the Hubble distance, $D_M(z)=c\int_0^z dz'/H(z')$ denotes the transverse comoving distance, $D_V(z)=\left[zD_M^2(z)D_H(z)\right]^{1/3}$ denotes the volume-averaged distance, and $r_d$ is the sound horizon at the baryon drag epoch.
\end{itemize}

We also perform a Bayesian model comparison between the CKG and $\Lambda$CDM models. The Bayesian evidence, $\mathcal{Z}$, is computed using the {\tt MCEvidence} package~\cite{heavens2017marginal,heavens2017no}~\href{https://github.com/yabebalFantaye/MCEvidence.git}{\faGithub}, accessed through the Cobaya interface implemented in the {\tt wgcosmo} repository~\href{https://github.com/williamgiare/wgcosmo.git}{\faGithub}. The relative performance of the two models is assessed via the logarithmic Bayes factor, $\ln B = \ln \mathcal{Z}_{\Lambda\mathrm{CDM}} - \ln \mathcal{Z}_{\rm CKG}.$ With this convention, positive values of $\ln B$ indicate a preference for the $\Lambda$CDM model over CKG, whereas negative values favor the CKG model over $\Lambda$CDM. The strength of the evidence is interpreted according to the revised Jeffreys scale \cite{kass1995bayes,trotta2008bayes}: $\ln B<1$ corresponds to inconclusive evidence, $1\leq\ln B<2.5$ to weak evidence, $2.5\leq\ln B<5$ to moderate evidence, $5\leq\ln B<10$ to strong evidence, and $\ln B\geq10$ to decisive evidence, in favor of whichever model yields the larger evidence. The priors adopted for the $\Lambda$CDM and CKG parameters are summarized in Table~\ref{tab_1}.

\begin{figure*}
\begin{subfigure}{.49\textwidth}
\includegraphics[width=\linewidth]{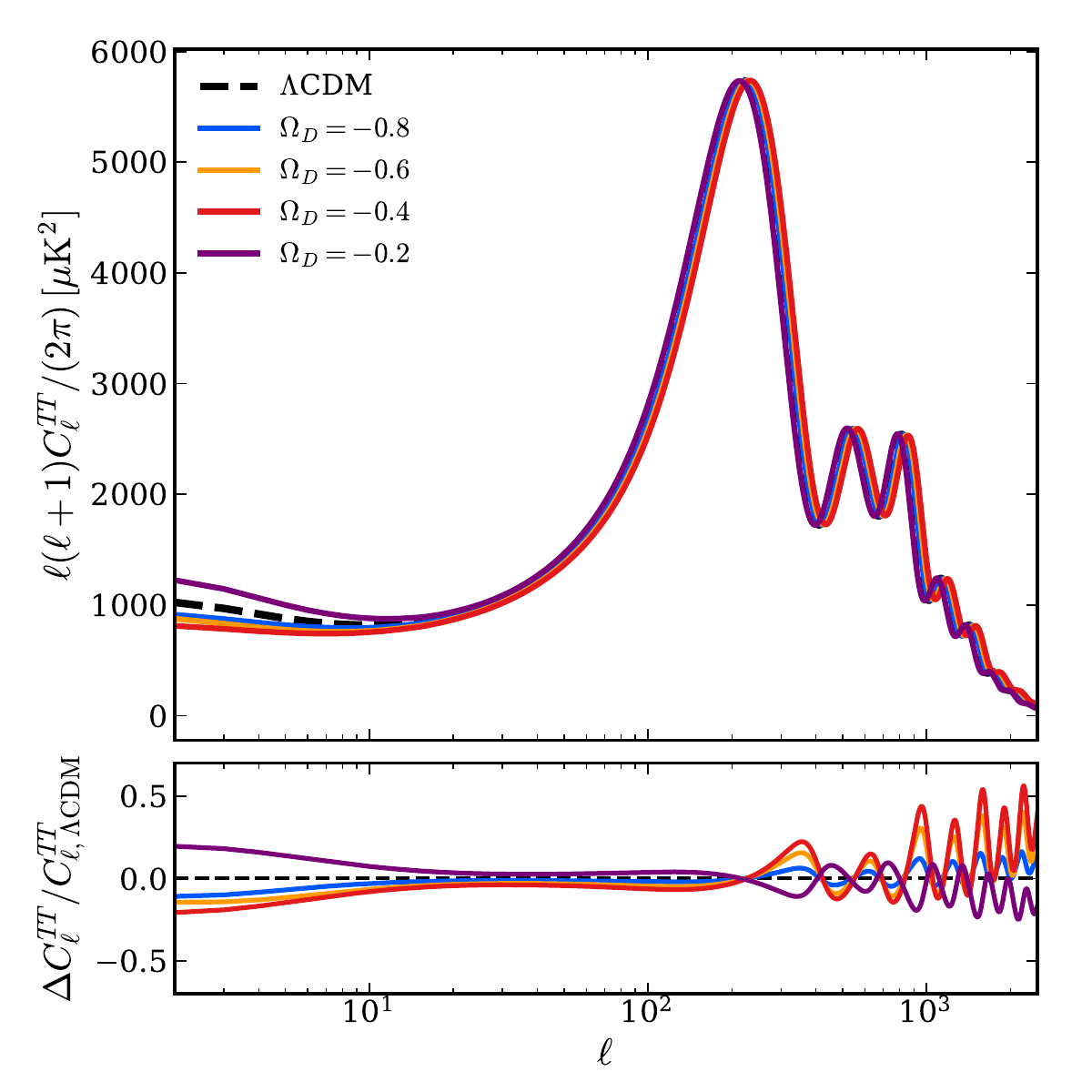}
\end{subfigure}
\hfil
\begin{subfigure}{.49\textwidth}
\includegraphics[width=\linewidth]{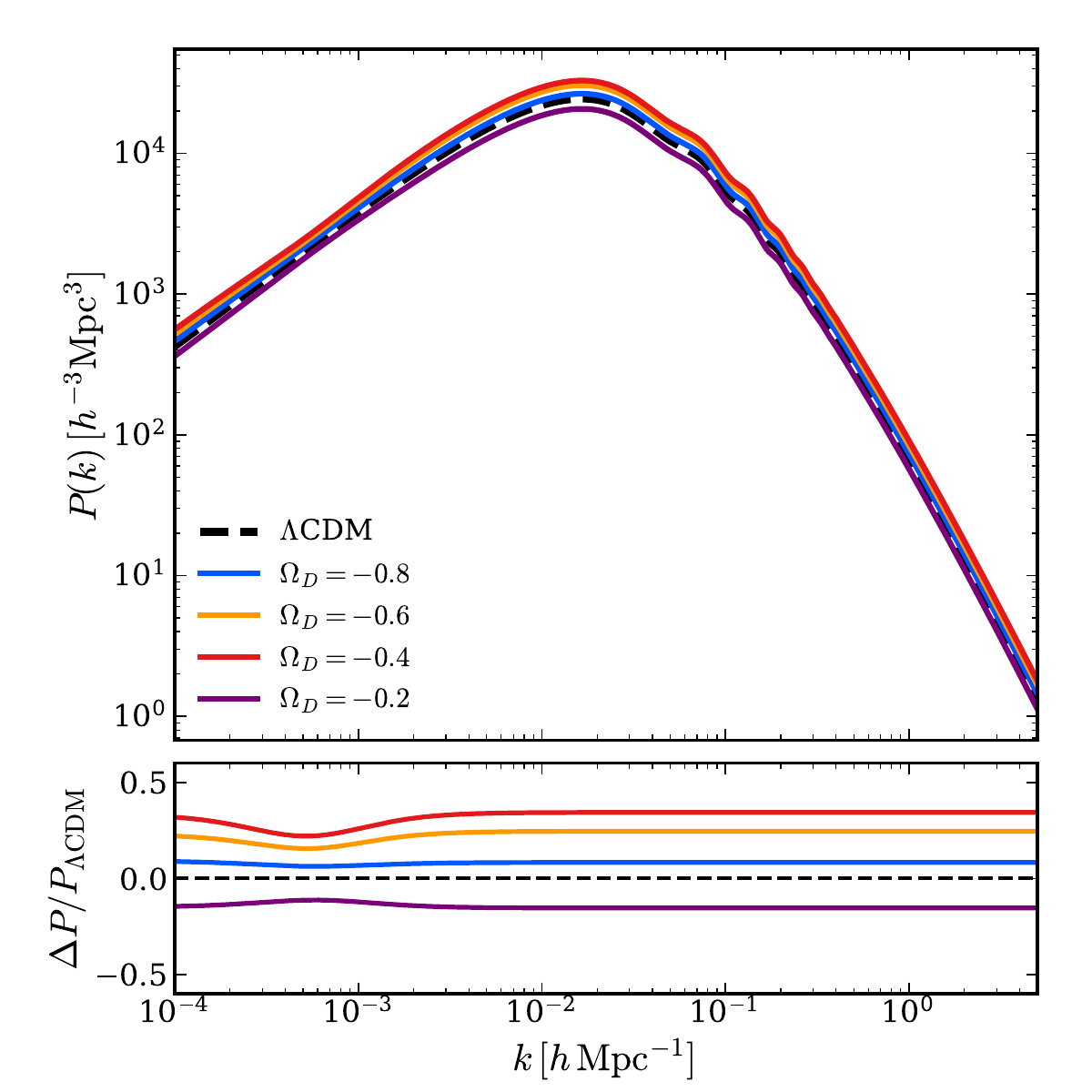}
\end{subfigure}
\caption{This figure shows the CMB temperature power spectrum, $C_\ell^{TT}$ (left panel), together with the corresponding fractional difference, $\Delta C_\ell^{TT}/C_{\ell,\Lambda{\rm CDM}}^{TT}$ (lower-left panel). The right panel shows the linear matter power spectrum, $P(k)$, and its corresponding fractional difference, $\Delta P/P_{\Lambda{\rm CDM}}$ (lower-right panel). The black dashed curve represents the $\Lambda$CDM model, while the colored curves show the predictions of the CKG model for different values of the parameter $\Omega_D$, with all other cosmological parameters fixed to their fiducial values.}\label{fig_1}
\end{figure*}
\begin{figure}
\centering
\includegraphics[scale=0.43]{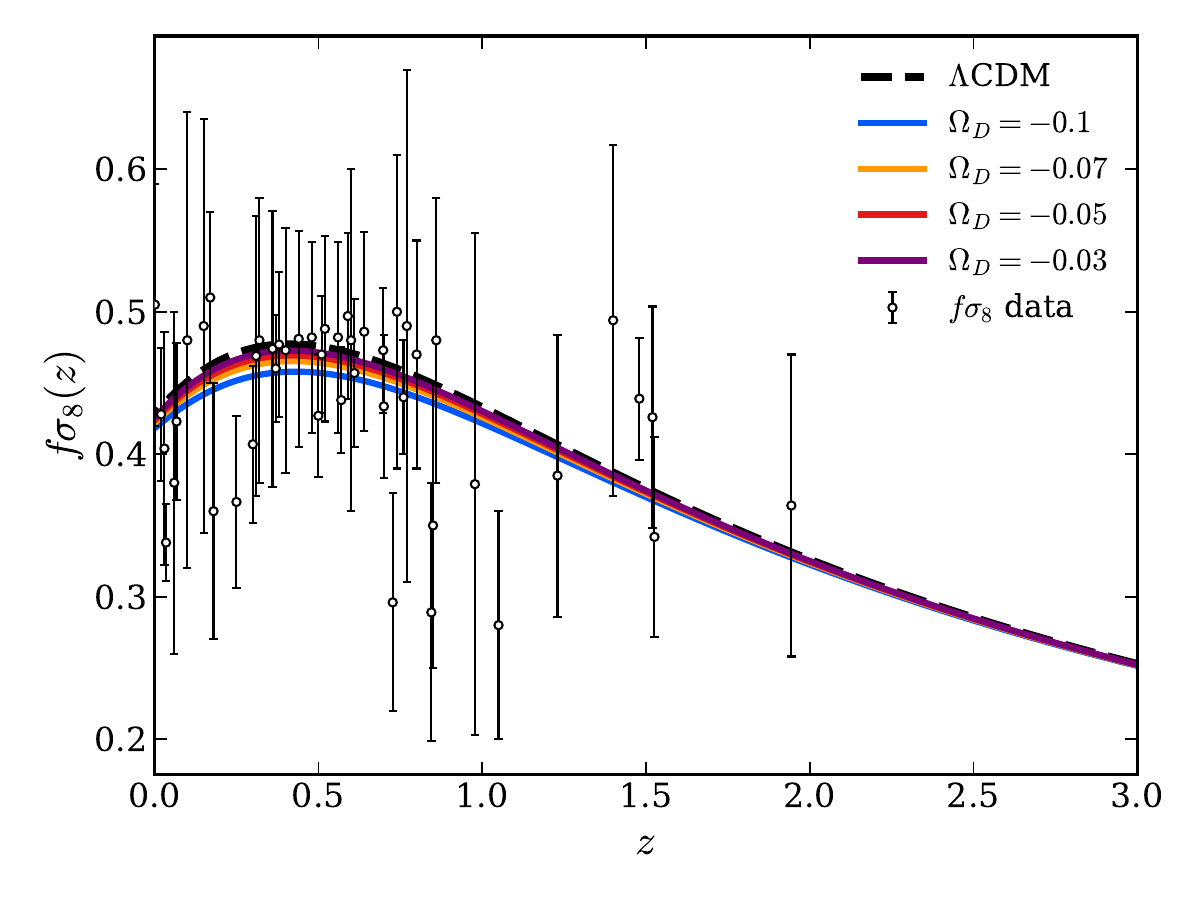}
\caption{This figure shows the evolution of the growth-rate observable $f\sigma_8(z)$ as a function of redshift for different values of the CKG model parameter $\Omega_D$, together with the observational measurements. The black dashed curve corresponds to the $\Lambda$CDM model, while the coloured curves show the predictions of the CKG model for different values of $\Omega_D$, with all other cosmological parameters fixed to their fiducial values.}\label{fig_2}
\end{figure}

Before constraining the CKG model with observational data, we first investigate its impact on the CMB temperature power spectrum (TT), the linear matter power spectrum ($P(k)$), and the growth-rate observable ($f\sigma_8$). These three observables provide complementary information on the expansion history of the Universe and the growth of cosmic structures, making them particularly useful for showing the cosmological effects of the CKG model. Fig.~\ref{fig_1} shows the TT (left panels) and the linear matter power spectrum $P(k)$ (right panels) for several values of the CKG model parameter $\Omega_D$. The upper panels show the absolute spectra, while the lower panels show the corresponding fractional differences relative to the $\Lambda$CDM model. Fig.~\ref{fig_2} shows the corresponding evolution of the growth-rate observable $f\sigma_8(z)$, with the observational $f\sigma_8$ measurements overlaid for comparison. The error bars are taken from Table~1 of Ref.~\cite{benisty2021quantifying}.

The left panels show that the CKG model has only a small effect on the TT power spectrum. The largest changes occur at low multipoles ($\ell\lesssim30$), where larger values of $\Omega_D$ increase the temperature anisotropies, while more negative values reduce them relative to the $\Lambda$CDM model. On the other hand, the positions of the acoustic peaks remain almost unchanged, indicating that the CKG model has little effect on the acoustic scale at recombination. Only small changes are seen in the amplitudes of the acoustic peaks, with the first peak becoming slightly higher for larger values of $\Omega_D$. This behavior is more clearly shown in the lower-left panel, where the fractional difference remains close to zero over most multipoles and exhibits small oscillations at high $\ell$. These oscillations follow the acoustic peak structure and indicate slight changes in the peak amplitudes, while their positions remain essentially unchanged.

The right panels show the corresponding linear matter power spectrum. Compared to the TT power spectrum, the CKG model has a more noticeable effect on the growth of matter perturbations. Larger values of $\Omega_D$ increase the matter power spectrum over the entire range of scales considered, while more negative values suppress it relative to the $\Lambda$CDM prediction. Despite these changes in amplitude, the overall shape of the spectrum remains nearly unchanged, indicating that the CKG model mainly affects the overall growth of matter rather than introducing a strong scale dependence. This behavior is more clearly seen in the lower-right panel, where the fractional difference is almost constant over most of the $k$ range, with only a small deviation on the largest scales. Overall, the CKG parameter primarily changes the amplitude of the linear matter power spectrum, while leaving its scale dependence largely unchanged.

The CKG model has only a small effect on the evolution of the growth-rate observable $f\sigma_8(z)$. Compared to the $\Lambda$CDM model,  CKG  predicts a slightly lower growth rate over the entire redshift range considered. The suppression becomes more noticeable for more negative values of $\Omega_D$, while models with $\Omega_D$ closer to zero are almost indistinguishable from the $\Lambda$CDM prediction. The largest differences are found at intermediate redshifts, whereas the curves become increasingly similar at higher redshifts. This suggests that the CKG parameter mainly affects the late-time growth of matter perturbations, with only a negligible impact on the early Universe. The observational $f\sigma_8$ measurements are consistent with both the $\Lambda$CDM and CKG predictions within the current uncertainties, indicating that the CKG model remains compatible with existing measurements of the growth of cosmic structures.

{
\renewcommand{\arraystretch}{1}
\begin{table}[t] 
    \centering
    \begin{tabular}{|lll|}
    \hline
    parametrization & parameter & prior\\  
    \hline 
    $\mathbf{\Lambda}$\textbf{CDM} & $\Omega_\mathrm{cdm}h^2$ & $\mathcal{U}[0.001, 0.99]$ \\   
    & $\Omega_\mathrm{b}h^{2}$ & $\mathcal{U}[0.005, 0.1]$ \\
    & $H_0$ & $\mathcal{U}[20.0, 100.0]$ \\
    & $\ln(10^{10} A_\mathrm{s})$ & $\mathcal{U}[1.61, 3.91]$ \\
    & $n_\mathrm{s}$ & $\mathcal{U}[0.8, 1.2]$ \\
    & $\tau$ & $\mathcal{U}[0.01, 0.8]$ \\
    \hline 
    \textbf{CKG} & $\Omega_D$ & $\mathcal{U}[-1, 0.02]$ \\
    \hline
    \end{tabular}
    \caption{ Parameters and priors used in the analysis. Here $\mathcal{U}[{\rm min, max}]$ denotes a uniform prior over the specified range.}\label{tab_1}
\end{table}
}


\begin{figure*}
\centering
\includegraphics[scale=0.40]{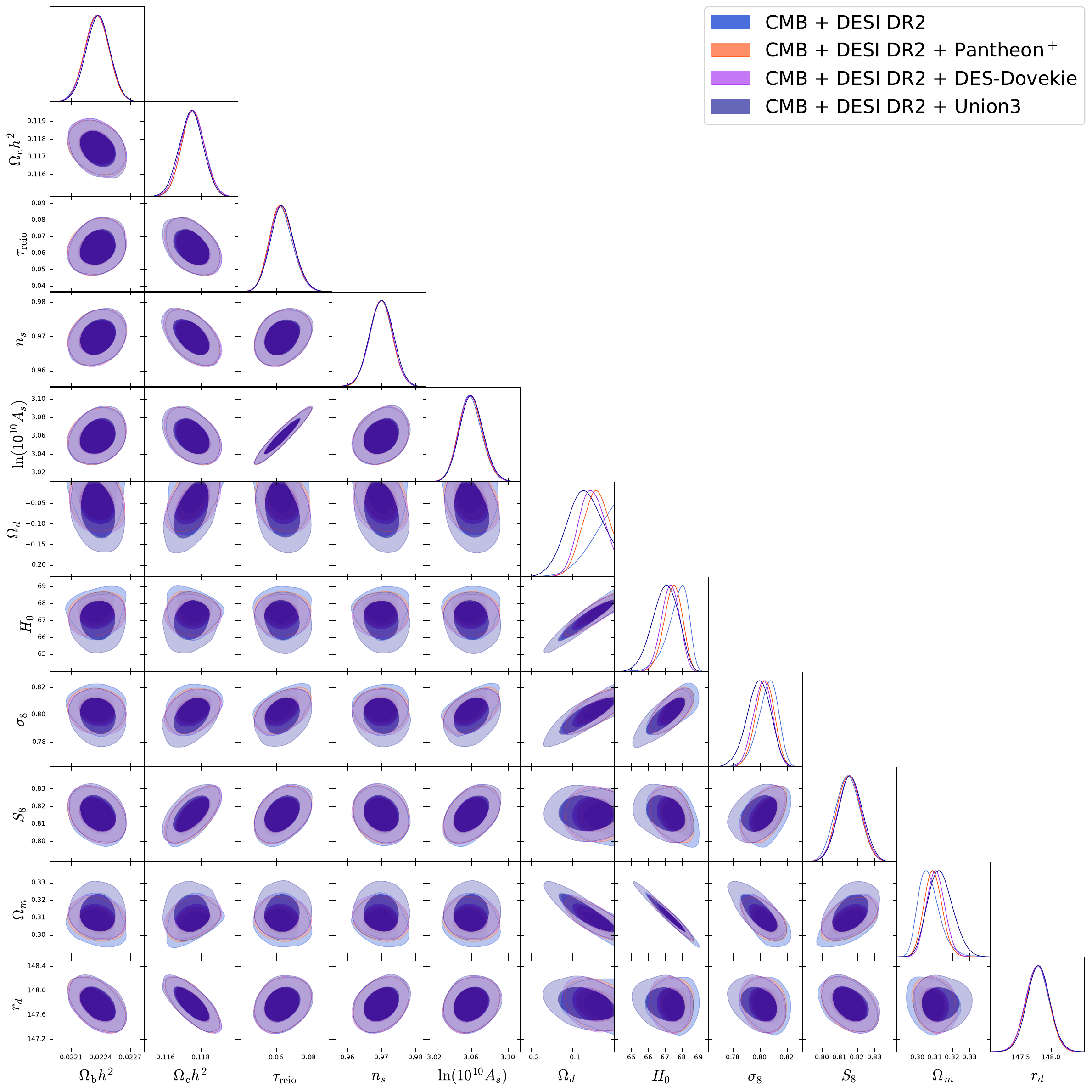}
\caption{The figure shows the corner plot of the CKG model at 68\% ($1\sigma$) and 95\% ($2\sigma$) confidence levels using DESI DR2 combined with CMB and SNe~Ia datasets (Pantheon$+$, DES-Dovekie, and Union3), shown as superimposed contours for the different dataset combinations.}\label{fig_3}
\end{figure*}

\begin{figure*}
\centering
\includegraphics[scale=0.44]{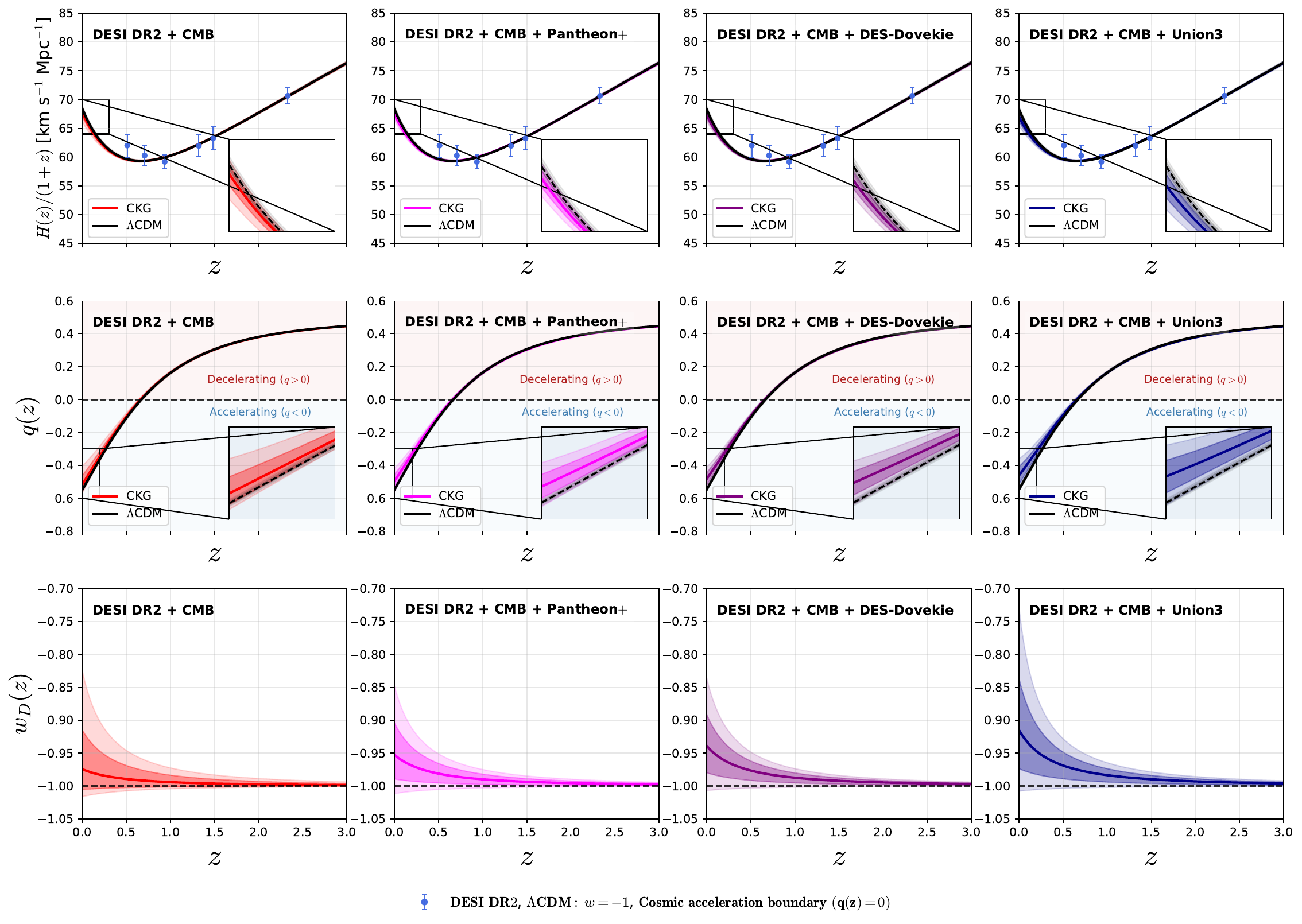}
\caption{The figure shows the evolution of $H(z)/(1+z)$ (first row), $q(z)$ (second row), $w(z)$ (third row), and $f_{\mathrm{DE}}(z)$ (fourth row) for the CKG and $\Lambda$CDM models as functions of $z$. The solid lines represent the mean predictions, while the light and dark shaded regions correspond to the $1\sigma$ and $2\sigma$ confidence intervals, respectively.}\label{fig_4}
\end{figure*}

\begin{table*}
\centering
\resizebox{\textwidth}{!}{%
\begin{tabular}{l@{\hspace{10pt}}c@{\hspace{10pt}}c@{\hspace{10pt}}c@{\hspace{10pt}}c}
\hline
\textbf{Parameters} 
& \textbf{CMB + DESI DR2} 
& \textbf{CMB + DESI DR2 + Pantheon$^+$} 
& \textbf{CMB + DESI DR2 + DES-Dovekie} 
& \textbf{CMB + DESI DR2 + Union3} \\
\hline
$H_0 (\mathrm{km\,s^{-1}\,Mpc^{-1}})$ 
& $69.64_{-0.93}^{+0.75}$ 
& $68.45_{-0.45}^{+0.34}$ 
& $68.34_{-0.40}^{+0.31}$ 
& $68.58_{-0.55}^{+0.37}$  \\

$\Omega_{\mathrm{\text{cdm}}}h^2$ 
& $0.11756 \pm 0.00065$ 
& $0.11753 \pm 0.00064$
& $0.11752 \pm 0.00067$
& $0.11744 \pm 0.00067$ \\

$\Omega_{\mathrm{\text{b}}}h^2$ 
& $0.02236 \pm 0.00012$ 
& $0.02236 \pm 0.00012$
& $0.02236 \pm 0.00012$
& $0.02237 \pm 0.00012$ \\

$\tau_{\mathrm{reio}}$ 
& $0.0631_{-0.0077}^{+0.0064}$
& $0.0632_{-0.0076}^{+0.0067}$
& $0.0634_{-0.0075}^{+0.0066}$
& $0.0638_{-0.0075}^{+0.0067}$ \\

$\ln\bigl(10^{10}\,A_{\mathrm{s}}\bigr)$ 
& $3.059_{-0.013}^{+0.012}$
& $3.059_{-0.013}^{+0.012}$
& $3.059 \pm 0.013$
& $3.060 \pm 0.013$ \\

$100\,\theta_{\mathrm{MC}}$ 
& $1.04218_{-0.00083}^{+0.00056}$ 
& $1.04129_{-0.00037}^{+0.00027}$ 
& $1.04218_{-0.00032}^{+0.00026}$ 
& $1.04138_{-0.00042}^{+0.00029}$  \\

$n_{\mathrm{s}}$ 
& $0.9698 \pm 0.0033$ 
& $0.9697 \pm 0.0033$
& $0.9698 \pm 0.0034$
& $0.9699 \pm 0.0035$ \\

$\Omega_d$ 
& $>-0.0413$ 
& $-0.043 \pm 0.032$
& $-0.053 \pm 0.031$
& $-0.069 \pm 0.043$ \\
\hline

$\Omega_{\mathrm{m}}$ 
& $0.3067_{-0.0073}^{+0.0046}$
& $0.3088 \pm 0.0052$
& $0.3103 \pm 0.0051$
& $0.3131_{-0.0078}^{+0.0063}$ \\

$\sigma_{8}$ 
& $0.8056_{-0.0062}^{+0.0086}$ 
& $0.8034_{-0.0065}^{+0.0072}$
& $0.8020 \pm 0.0069$ 
& $0.7989_{-0.0081}^{+0.0094}$  \\

$S_{8}$ 
& $0.8144 \pm 0.0070$ 
& $0.8151 \pm 0.0067$ 
& $0.8156 \pm 0.0067$ 
& $0.8160 \pm0.0068$   \\

$r_d\text{(Mpc)}$ 
& $147.77 \pm 0.19$ 
& $147.77 \pm 0.19$
& $147.77 \pm 0.19$
& $147.79 \pm 0.19$ \\

\hline
$\Delta{\chi^{2}_{\text{MAP}}}$
& 1.13
& -1.20
& -3.01
& -2.49
\\

$\ln B_{\Lambda \rm CDM,\; CKG}$
& $-5.53$
& $-6.08$
& $-6.84$
& $-6.61$
\\

\hline
\end{tabular}
}
\caption{This table shows the constraints on the AS parameters, presenting the mean values and their uncertainties at the 68\% ($1\sigma$) confidence level, obtained using DESI DR2 combined with CMB and SNe~Ia datasets (Pantheon$^+$, DES-Dovekie, and Union3).}
\label{tab_2}
\end{table*}

\section{Results}\label{sec_4}
Fig.~\ref{fig_3} presents the marginalized posterior distributions of the CKG model parameters at the 68\% ($1\sigma$) and 95\% ($2\sigma$) confidence levels, obtained using the combinations CMB + DESI DR2, CMB + DESI DR2 + Pantheon$+$, CMB + DESI DR2 + DES-Dovekie, and CMB + DESI DR2 + Union3. The off-diagonal panels shows the two-dimensional (2D) joint marginalized constraints for different pairs of cosmological parameters, whereas the diagonal panels show the corresponding 1D marginalized posterior distributions.

We begin by examining the status of the $H_0$ tension within the CKG model using the latest DESI DR2 observations. A key point to keep in mind is the well-established inverse relationship between the Hubble constant, $H_0$, and the sound horizon at the drag epoch, $r_d$, as discussed by \cite{knox2020hubble}. In simple terms, increasing the inferred value of $H_0$ requires a corresponding decrease in $r_d$, and vice versa. Consequently, matching the local Riess determination, $H_0=(73.04\pm1.04),\mathrm{km,s^{-1},Mpc^{-1}}$ \cite{riess2022comprehensive,Murakami:2023xuy,Breuval:2024lsv}, would require the sound horizon to be reduced by approximately $7\%$.

In the case of the CKG model, before discussing the implications of the $H_0$ tension, it is instructive to examine the physical role of the CKG parameter, $\Omega_D$. As can be seen from Eq.~(\ref{HSQ}), the CKG contribution enters the Friedmann equation as $\Omega_D/(1+z)^2$. This redshift dependence is fundamentally different from those of radiation, matter, and spatial curvature, which scale as $(1+z)^4$, $(1+z)^3$, and $(1+z)^2$, respectively. Consequently, during the pre-recombination epoch ($z\gg1$), the contribution of the $\Omega_D$ term is strongly suppressed by the factor $(1+z)^{-2}$ and is therefore negligible compared with the radiation- and matter-dominated components. As the Universe expands into the post-recombination epoch and the redshift decreases, this suppression gradually weakens, allowing the CKG contribution to become increasingly important. Therefore, the parameter $\Omega_D$ modifies only the post-recombination expansion history, while leaving the pre-recombination evolution essentially unchanged.

This behavior has an important implication for the $H_0$ tension in the CKG model. Since the CKG parameter $\Omega_D$ modifies only the post-recombination expansion history of the Universe, it leaves the pre-recombination evolution essentially unchanged. Consequently, the sound horizon at the drag epoch, $r_d$, remains essentially unchanged. Although the CKG model predicts a relatively higher value of $H_0$, reducing the discrepancy with the Riess measurement to approximately $2.65\sigma$ for the CMB + DESI DR2 combination, it yields a nearly unchanged sound horizon of $r_d=147.77\pm0.19~\mathrm{Mpc}$. Similarly, after including the SNe~Ia datasets, the inferred value of $H_0$ shifts toward lower values, resulting in tensions of approximately $4.19\sigma$, $4.33\sigma$, and $4.04\sigma$ for the CMB + DESI DR2 + Pantheon$^+$, CMB + DESI DR2 + DES-Dovekie, and CMB + DESI DR2 + Union3 combinations, respectively. At the same time, the corresponding sound horizon remains nearly unchanged, with $r_d=147.77\pm0.19~\mathrm{Mpc}$ for both the Pantheon$^+$ and DES-Dovekie datasets and $r_d=147.79\pm0.19~\mathrm{Mpc}$ for the Union3 dataset. These results show the implications for the $H_0$ tension and explain why the CKG model cannot resolve the $H_0$ tension, as it leaves the sound horizon essentially unaffected and, therefore, is unable to provide a complete solution of the $H_0$ tension.

An alternative approach to reducing the sound horizon, $r_d$, is to modify the expansion history of the Universe during the pre-recombination epoch, approximately between $z\sim10^4$ and $z\sim10^3$. Such modifications generally require an increase in the physical matter density, as realized in Early Dark Energy (EDE) scenarios~\cite{poulin2019early}. Although EDE models can successfully fit the combined BAO and CMB observations and alleviate the $H_0$ tension, they generally provide a poorer fit to SNe~Ia data and tend to exacerbate the existing $S_8$ tension~\cite{chaussidon2025early}.

We further examine the $S_8 \equiv \sigma_8\left(\Omega_m/0.3\right)^{0.5}$ parameter within the CKG model, as it characterizes the amplitude of matter clustering and plays a central role in the current $S_8$ tension. Recent weak-lensing observations, including DES~Y6 ($S_8 = 0.794^{+0.009}_{-0.012}$)~\cite{DES:2026fyc} and KiDS-Legacy (Shear) ($S_8 = 0.815^{+0.016}_{-0.021}$)~\cite{wright2025kids}, prefer lower values than those inferred from the combined Planck + ACT + SPT CMB analysis, which reports $S_8 = 0.836^{+0.012}_{-0.013}$~\cite{DES:2026fyc}. These values correspond to tensions of about $2.4\sigma$ for DES~Y6 and $0.9\sigma$ for KiDS-Legacy (Shear). A comprehensive compilation of $S_8$ measurements and their corresponding tensions relative to the combined Planck + ACT + SPT CMB constraint is provided in Table~1 of Ref.~\cite{pantos2026status}.

In our analysis, we compare the inferred $S_8$ values with the recent weak-lensing measurements from KiDS-Legacy (Shear) and DES~Y6 to assess the consistency of the CKG model with current large-scale structure observations. Relative to the KiDS-Legacy (Shear) measurement, the CMB + DESI DR2, CMB + DESI DR2 + Pantheon$^+$, CMB + DESI DR2 + DES-Dovekie, and CMB + DESI DR2 + Union3 dataset combinations shows tensions of $0.03\sigma$, $0.01\sigma$, $0.03\sigma$, and $0.05\sigma$, respectively. The corresponding tensions with the DES~Y6 measurement are $1.62\sigma$, $1.69\sigma$, $1.73\sigma$, and $1.76\sigma$, respectively. These results indicate that the CKG model is in good agreement with the KiDS-Legacy (Shear) measurement and remains compatible with the DES~Y6 weak-lensing constraints, showing no significant $S_8$ tension.

The parameter $\Omega_D$ quantifies the present-day fractional energy density of the geometric dark-energy component arising from the conformal Killing symmetry. Its sign and magnitude determine the contribution of this geometric term to the cosmic energy budget and, consequently, influence the dynamical behavior of dark energy. The best-fit values of $\Omega_D$ are consistently negative for all dataset combinations. This implies that the effective dark energy behaves as a quintessence-like component ($w_D>-1$) rather than a phantom one ($w_D<-1$). Therefore, within the conformal Killing gravity framework, current observations favor a dark-energy sector that evolves on the quintessence side of the phantom divide, although the preference remains statistically insignificant.

To further investigate the future evolution predicted by the CKG model, we solve Eq.~(\ref{ZCRIT}) for the critical redshift $z_c$ using the full MCMC posterior distributions. We obtain
\[
z_c =
\begin{cases}
-0.772^{+0.103}_{-0.102}, & \text{CMB + DESI DR2},\\
-0.747^{+0.068}_{-0.088}, & \text{CMB + DESI DR2 + Pantheon$^+$},\\
-0.728^{+0.059}_{-0.084}, & \text{CMB + DESI DR2 + DES-Dovekie},\\
-0.690^{+0.065}_{-0.098}, & \text{CMB + DESI DR2 + Union3}.
\end{cases}
\]

For all dataset combinations, the inferred critical redshift satisfies $-1<z_c<0$, implying that the Hubble expansion rate vanishes only in the future. Consequently, the best-fit CKG cosmology favors a scenario in which the current accelerated expansion eventually slows to a halt at a finite future epoch before the formal singular limit at $z=-1$ is reached. Therefore, the cosmological evolution remains physically well-defined up to the turning point, beyond which the standard expanding background solution is no longer applicable.

The maximum-likelihood comparison indicates a preference for the CKG model over $\Lambda$CDM. For the CMB + DESI DR2 dataset, we obtain $\Delta\chi^{2}_{\rm MAP}=1.13$, indicating a marginally better best fit for $\Lambda$CDM. However, after including Pantheon$^+$, DES-Dovekie, and Union3, the values become $\Delta\chi^{2}_{\rm MAP}=-1.20$, $-3.01$, and $-2.49$, respectively, showing that the CKG model provides a better best fit than $\Lambda$CDM. In contrast, the logarithm of the Bayes factor consistently favors the CKG model for all dataset combinations, with $\ln B_{\Lambda \rm CDM,\;CKG}=-5.53$, $-6.08$, $-6.84$, and $-6.61$, corresponding to strong Bayesian evidence in favor of the CKG model according to the Jeffreys scale. The agreement between the minimized $\chi^{2}$ and the Bayesian evidence, particularly for the dataset combinations including Type Ia supernova observations, indicates that the CKG model not only provides an improved best-fit description of the data but is also preferred when the full prior-weighted parameter space is taken into account.

In Fig.~\ref{fig_4}, we present the reconstructed evolution of $H(z)/(1+z)$ (first row), $q(z)$ (second row), and $w(z)$ (third row) as functions of redshift for the CKG and $\Lambda$CDM models. In each panel, the solid curves represent the mean reconstructed evolution, while the darker and lighter shaded regions correspond to the $1\sigma$ and $2\sigma$ confidence intervals, respectively. The first row shows the reconstructed evolution of $H(z)/(1+z)$. For the CMB + DESI DR2 dataset, the CKG model predicts a slightly lower expansion history than $\Lambda$CDM around the minimum of $H(z)/(1+z)$, although the two reconstructions remain fully consistent within their $1\sigma$ confidence regions. The inclusion of the Pantheon$^+$, DES-Dovekie, and Union3 supernova samples further reduces the differences between the two models, yielding nearly identical expansion histories over the entire redshift range. The residual deviations are only visible in the zoomed-in panels, where the CKG model shows a marginally lower expansion rate than $\Lambda$CDM at intermediate redshifts. Overall, both models provide an equally good description of the current expansion history of the Universe.

The second row presents the reconstructed evolution of the deceleration parameter, $q(z)$, for the CKG and $\Lambda$CDM models. Both models shows the expected transition from a decelerating ($q>0$) to an accelerating ($q<0$) Universe at similar redshifts. Compared to $\Lambda$CDM, the CKG model predicts a slightly less negative present-day deceleration parameter, with the differences becoming more noticeable after the inclusion of the Pantheon$^+$, DES-Dovekie, and Union3 supernova samples, as highlighted in the zoomed-in panels. Nevertheless, the reconstructed $q(z)$ curves remain consistent within the corresponding $1\sigma$ confidence regions. Overall, both models predict a present-day accelerated expansion of the Universe, in agreement with current astronomical observations.

The third row presents the reconstructed evolution of the dark energy equation of state, $w_D(z)$, for the CKG model, while the $\Lambda$CDM model corresponds to the constant value $w=-1$. For all four dataset combinations, the reconstructed equation of state remains entirely in the quintessence regime, with $w_D(z)>-1$ throughout the redshift range considered, and no evidence for a crossing of the phantom divide is found~\cite{Cai:2025mas}. Instead, the equation of state evolves smoothly from a mild deviation from $\Lambda$CDM at the present epoch towards $w=-1$ with increasing redshift, indicating that any dynamical effects are confined to the late Universe. Among the dataset combinations, the CMB + DESI DR2 reconstruction lies closest to the cosmological constant, with $w_D(0)\simeq-0.98$. The inclusion of supernova observations favors progressively larger departures from $\Lambda$CDM, with the Pantheon$+$, DES-Dovekie, and Union3 compilations yielding $w_D(0)\simeq-0.971$, $-0.965$, and $-0.954$, respectively. Despite these differences at low redshift, all reconstructions rapidly converge towards $w=-1$ by $z\gtrsim1$, demonstrating that the CKG model approaches the standard cosmological constant behavior at earlier cosmic times.

\section{Perspectives and Conclusion}
\label{sec_5}

In this work, we have investigated the cosmological implications of the CKG, in which the dark energy component arises directly from the conformal Killing symmetry of the RW space-time. Within this framework, the divergence-free conformal Killing tensor behaves as an effective perfect fluid whose energy density, pressure, and equation of state are completely determined by the geometry of the space-time. Consequently, the evolution of dark energy is not introduced through an ad hoc parametrization but follows directly from the geometric properties of the cosmological background. The resulting model extends the standard $\Lambda$CDM cosmology through a single additional parameter, $\Omega_D$, while naturally recovering the $\Lambda$CDM limit when $\Omega_D=0$.

To test the viability of this geometric framework, we performed a comprehensive Bayesian parameter estimation using the latest cosmological observations, including the Planck PR4 (NPIPE) CMB temperature, polarization and lensing measurements together with ACT DR6 CMB lensing, DESI DR2 BAO observations, and the Pantheon$+$, DES-Dovekie, and Union3 Type Ia supernova compilations. These datasets provide stringent constraints on the late-time expansion history and allow a detailed comparison between the CKG model and the standard $\Lambda$CDM scenario.

Our analysis shows that the geometric dark-energy parameter remains tightly constrained by current observations and consistently favors a negative value, corresponding to a quintessence-like dark energy equation of state with $w_D(z)>-1$. No evidence for a crossing of the phantom divide is found, and the reconstructed equation of state smoothly approaches the cosmological constant value, $w=-1$, at higher redshifts. Similarly, the reconstructed expansion history and deceleration parameter remain fully consistent with the observed transition from a matter-dominated decelerating Universe to the present epoch of accelerated expansion, while differing only slightly from the predictions of the standard $\Lambda$CDM model at late times. An additional prediction of the CKG model is the existence of a future critical redshift, $z_c$, at which the Hubble expansion rate vanishes. For all dataset combinations, we find $z_c$ in the range $-0.8 \lesssim z_c \lesssim -0.7$, indicating a future turning point of the cosmic expansion before the formal singular limit at $z=-1$ is reached.

The geometric nature of the CKG  also provides a clear physical interpretation of its cosmological effects. Since the additional term scales as $(1+z)^{-2}$, its contribution is strongly suppressed during the radiation- and matter-dominated epochs and becomes relevant only after recombination. Consequently, the sound horizon at the baryon drag epoch remains essentially unchanged, implying that the model cannot provide a complete resolution of the current $H_0$ tension. Nevertheless, the predicted values of $S_8$ remain in good agreement with current weak-lensing observations, exhibiting tensions of less than $0.1\sigma$ with KiDS-Legacy (Shear) and less than $1.8\sigma$ with DES~Y6, indicating that the model is fully consistent with current weak-lensing measurements.

From the statistical point of view, the CKG model performs remarkably well. While the maximum-likelihood analysis shows comparable or improved fits relative to $\Lambda$CDM, particularly after the inclusion of Type Ia supernova observations, the Bayesian model comparison consistently yields strong evidence in favor of the CKG scenario according to the revised Jeffreys scale. These results indicate that the conformal Killing framework is not only theoretically well motivated but it is also observationally competitive with, and statistically preferred over, the standard cosmological model when confronted with current cosmological data.

Overall, our results demonstrate that CKG provides a simple, physically motivated, and observationally successful description of late-time cosmic acceleration. Rather than postulating an empirical dark-energy parametrization, the accelerated expansion emerges naturally from a fundamental geometric symmetry of the space-time itself, in the same line of previous results supposing a geometric origin of dark energy \cite{Capozziello:2002rd,Nojiri:2010wj,Nojiri:2017ncd,Clifton:2011jh, DeFelice:2010aj}. Future high-precision cosmological observations from Euclid, the Vera C. Rubin Observatory Legacy Survey of Space and Time (LSST), the Nancy Grace Roman Space Telescope, DESI DR3, and CMB-S4 will substantially improve the precision of cosmological measurements and offer stringent tests of this geometric framework. These next-generation surveys will determine whether the subtle departures from $\Lambda$CDM predicted by the conformal Killing cosmology can be detected and will further clarify whether the origin of dark energy is fundamentally geometric in nature.

\section*{Acknowledgements}
SC acknowledges the Istituto Nazionale di Fisica Nucleare (INFN) Sez. di Napoli,  Iniziative Specifiche QGSKY and MoonLight-2  and the Istituto Nazionale di Alta Matematica (INdAM), gruppo GNFM, for the support. This paper is based upon work from COST Action CA21136 -- Addressing observational tensions in cosmology with systematics and fundamental physics (CosmoVerse), supported by COST (European Cooperation in Science and Technology).

\bibliographystyle{elsarticle-num}
\bibliography{mybib.bib}

\end{document}